\documentclass[aps,pra,reprint,superscriptaddress]{revtex4-2}

\usepackage{amsmath,amsfonts,amssymb}
\usepackage{xcolor,graphicx}
\usepackage{braket}
\usepackage{float}
\usepackage[pdftex]{hyperref}
\hypersetup{
	colorlinks = true,
	linkcolor = blue,
	anchorcolor = blue,
	citecolor = red,
	filecolor = blue,
	urlcolor = blue}
\usepackage{subfig}

\begin{document}

\title{Exact solution of the master equation for interacting quantized fields at finite temperature decay}

\author{L. Hernández-Sánchez}
\email[e-mail: ]{leonardi1469@gmail.com}
\affiliation{Instituto Nacional de Astrofísica Óptica y Electrónica, Calle Luis Enrique Erro No. 1\\ Santa María Tonantzintla, Puebla, 72840, Mexico}
\author{I. A. Bocanegra-Garay}
\affiliation{Departamento de F\'isica Te\'orica, At\'omica y \'Optica and  Laboratory for Disruptive Interdisciplinary Science, Universidad de Valladolid, 47011 Valladolid, Spain}
\author{I. Ramos-Prieto}
\affiliation{Instituto Nacional de Astrofísica Óptica y Electrónica, Calle Luis Enrique Erro No. 1\\ Santa María Tonantzintla, Puebla, 72840, Mexico}
\author{F. Soto-Eguibar}
\affiliation{Instituto Nacional de Astrofísica Óptica y Electrónica, Calle Luis Enrique Erro No. 1\\ Santa María Tonantzintla, Puebla, 72840, Mexico}
\author{H.M. Moya-Cessa}
\affiliation{Instituto Nacional de Astrofísica Óptica y Electrónica, Calle Luis Enrique Erro No. 1\\ Santa María Tonantzintla, Puebla, 72840, Mexico}

\date{\today}

\begin{abstract}
We analyze the Markovian dynamics of a quantum system involving the interaction of two quantized fields at finite temperature decay. Utilizing superoperator techniques and applying two non-unitary transformations, we reformulate the Lindblad master equation into a von Neumann-like equation with an effective non-Hermitian Hamiltonian. Furthermore, an additional non-unitary transformation is employed to diagonalize this Hamiltonian, enabling us to derive an exact solution to the Lindblad master equation. This method provides a framework to calculate the evolution of any initial state in a fully quantum regime. As a specific example, we present the photon coincidence rates for two indistinguishable photons initially interacting within a cavity.
\end{abstract}
\maketitle

\section{Introduction}
The study of the dynamics and evolution of quantum systems under environmental influence presents a significant challenge for both experimental and theoretical physics. In this context, the Lindblad master equation emerges as a fundamental tool, offering a formal framework that accurately describes the Markovian dynamics of open quantum systems~\cite{Carmichael_Book, Breuer_Book, Manzano_2020, Gao_2024}. This equation not only captures the dissipation processes within the system but also incorporates thermal fluctuations, which introduce additional channels for loss and excitation. However, obtaining an exact analytical solution to the master equation remains a considerable challenge, especially in finite-temperature systems where the complexities of interactions and decoherence further complicate the pursuit of precise results~\cite{Prosen_2012, Torres_2014, Minganti_2020, Teuber_2020, Czerwinski_2022}. Consequently, numerical methods often become the primary tool for analyzing such systems, although they typically demand extensive computational resources.

Recent developments in cavity quantum electrodynamics have led to the derivation of effective master equations for quantum systems coupled to bosonic modes, resulting in a substantial reduction of the Hilbert space. While these equations are able to describe certain system behaviors, they often lack the property of complete positivity, which can lead to inaccurate predictions regarding system dynamics~\cite{Bezvershenko_2021, Jager_2022, Jager_2023}. This highlights the need for exact solutions to the Lindblad master equation, as only such solutions ensure physical consistency and accuracy in describing the behavior of open quantum systems. 

On the other hand, in the analysis of the evolution of one or more electromagnetic field modes within a cavity, a technique based on superoperators has been developed that, through non-unitary transformations, reformulates the Lindblad master equation into a more tractable form. This method simplifies the analysis by eliminating complex terms, such as quantum jumps, and facilitates the application of solutions to various initial conditions, including coherent and number states~\cite{Phoenix_1990, Arévalo_1995, Arevalo_1998, Hernandez_2023}. In this context, recent investigations have demonstrated that the Markovian dynamics of decaying coupled harmonic oscillators can be effectively described using a non-Hermitian Hamiltonian. This approach has garnered increasing interest due to the unique properties of these systems, particularly regarding parity-time ($\mathcal{PT}$) symmetry and exceptional points (EPs)~\cite{Bender_1998, Guo_2009, Lin_2011, Prosen_2012, Heiss_2012, Feng_2017, ElGanainy_2018, MohammadAli_2019, Arkhipov_2020, Tschernig_2022}. Although significant progress has been made in the zero-temperature regime, analyzing systems at finite temperatures remains a considerable challenge. In this work, we aim to extend this approach to the finite-temperature regime, exploring the possibility of obtaining analogous results under such conditions.

The structure of this article is as follows: in Section~\ref{Sec_II}, we introduce the theoretical model that forms the basis of our research. In Section~\ref{Sec_III}, we show how, by means of two non-unitary transformations, the Lindblad master equation for interacting quantized fields at finite temperature can be recast into a von Neumann-like equation with an effective non-Hermitian Hamiltonian—one of the key results of this work. Section~\ref{Sec_IV} presents the exact solution to the Lindblad master equation and includes an example illustrating the photon coincidence rates for two indistinguishable photons initially interacting within a cavity. Finally, in Section~\ref{Conclusions}, we summarize our conclusions.

\begin{figure}[h]
\includegraphics[width= \linewidth]{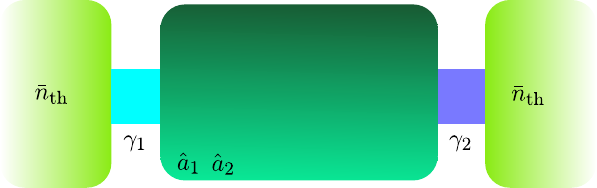}
\caption{The schematic representation illustrates two bosonic fields, $\hat{a}_1$ and $\hat{a}_2$, both sharing the same thermal distribution, characterized by the average thermal photon number $\bar{n}_{\text{th}}$, indicating that they are at the same temperature. However, the fields decay at different rates, denoted by $\gamma_1$ and $\gamma_2$, respectively, reflecting the distinct dissipation mechanisms that independently affect each field.}
\label{Fig1}
\end{figure}

\section{Theoretical model}\label{Sec_II}
We consider the interaction between two bosonic fields, denoted by $\hat{a}_1$ and $\hat{a}_2$, each subject to distinct Markovian decay processes characterized by the loss rates $\gamma_1$ and $\gamma_2$, respectively. The fields are also coupled to a thermal reservoir with an average thermal excitation number $\bar{n}{\text{th}} \geq 0$, as depicted in Fig.~\ref{Fig1}.  Due to the differences in their decay rates, each field experiences its own dissipation dynamics. The Markovian evolution of the density matrix $\hat{\rho}$, in the interaction picture, is governed by the Lindblad master equation\cite{Breuer_Book, Carmichael_Book, Manzano_2020}
\begin{equation}\label{master1}
\begin{split}
\frac{d\hat{\rho}}{dt} =& -\mathrm{i}g\hat{S}\hat{\rho}+\left(\bar{n}_{\text{th}}+1\right)\sum_{j=1}^2\gamma_j\left[2\hat{a}_j\hat{\rho}\hat{a}_j^\dagger-\{\hat{a}_j^\dagger\hat{a}_j,\hat{\rho}\}\right]\\	&+\bar{n}_{\text{th}}\sum_{j=1}^2\gamma_j\left[2\hat{a}_j^\dagger\hat{\rho}\hat{a}_j-\{\hat{a}_j\hat{a}_j^\dagger,\hat{\rho}\}\right],
\end{split}
\end{equation}
where $\hat{a}_j$ and $\hat{a}_j^\dagger$ are the annihilation and creation operators, respectively, for each mode $j = 1, 2$, and $g$ represents the coupling strength between the two bosonic modes~\cite{Louisell_1990, Vogel_2006, Fox_2006, Garrison_2008, Klimov_2009, Moya_Book, agarwal_2013, Gerry_Book}.

The right-hand side of Eq.~\eqref{master1} contains three primary contributions:
\begin{enumerate}
\item The first term, $-\mathrm{i}g\hat{S}\hat{\rho}$, describes the coherent interaction between the two bosonic modes. This interaction is represented by $\hat{S}\hat{\rho} = [\hat{a}_1\hat{a}_2^\dagger + \hat{a}_1^\dagger\hat{a}_2, \hat{\rho}]$, which models the exchange of excitations between the modes due to their coupling.
\item The second term accounts for the loss of excitations due to interaction with the thermal environment. The factor $\bar{n}_{\text{th}} + 1$ reflects the total number of excitations, including those induced by the thermal reservoir. The expression inside the summation models the dissipation of excitations through the annihilation operators $\hat{a}_j$.
\item The third term captures the gain of thermal excitations from the environment. Here, $\bar{n}_{\text{th}}$ represents the average number of thermal excitations at equilibrium, and the term in square brackets describes the gain of excitations through the creation operators $\hat{a}_j^\dagger$.
\end{enumerate}
Together, these terms provide a comprehensive description of the dynamics of the bosonic system, highlighting how both the mutual interaction and the thermal environment influence the dissipation and excitation of the modes.

\section{Finite temperature Lindblad master equation as a Von Neumann-like equation}\label{Sec_III}

In the study of open quantum systems, superoperators provide a useful framework for representing complex dynamics in a more accessible way. Unlike conventional operators in quantum mechanics, which act on state vectors, superoperators operate directly on the density matrix. This feature allows them to comprehensively capture both the coherent evolution and the dissipative effects described by the Lindblad master equation.

With this perspective, we introduce the following superoperators \cite{Phoenix_1990, Arévalo_1995, Arevalo_1998, Hernandez_2023}
\begin{equation}\label{superJ}
		\hat{J}_j^{(-)}\hat{\rho}:= 2\hat{a}_j\hat{\rho}\hat{a}_j^\dagger,\qquad
		\hat{J}_j^{(+)}\hat{\rho}:= 2\hat{a}_j^\dagger\hat{\rho}\hat{a}_j,
\end{equation}
and
\begin{equation}
	\begin{split}
		\hat{L}_j^{(-)}\hat{\rho}&:= \hat{a}_j^\dagger\hat{a}_j\hat{\rho}+\hat{\rho}\hat{a}_j^\dagger\hat{a}_j,\\
		\hat{L}_j^{(+)}\hat{\rho}&:= \hat{a}_j\hat{a}_j^\dagger\hat{\rho}+\hat{\rho}\hat{a}_j\hat{a}_j^\dagger = \left[\hat{L}_j^{(-)}+2\right]\hat{\rho},
	\end{split}
\end{equation}
with $j=1,2$.
From these definitions, we can derive the following relations
\begin{equation}
\begin{split}
\left[\hat{J}_j^{(-)},\hat{J}_j^{(+)}\right]\hat{\rho}&= 4\left(\hat{L}_j+1\right)\hat{\rho},\\
\left[\hat{L}_j,\hat{J}_{j}^{(\pm)}\right]\hat{\rho}&= \pm 2\hat{J}_j^{(\pm)}\hat{\rho},\\
\left[\hat{J}_1^{(\pm)}+\hat{J}_2^{(\pm)},\hat{S}\right]\hat{\rho}&= 0,
\end{split}
\end{equation}
where $\hat{L}_j\hat\rho\equiv \hat{L}_j^{(-)}\hat\rho$. The operators defined as $\hat A_0\hat\rho:=\frac{1}{2}(\hat{L}_j+1)\hat\rho$ and $\hat A^{(\pm)}\hat\rho:=\frac{1}{2} \hat J_j^{(\pm)}\hat\rho$, satisfy the commutation relations of the  $\mathfrak{su}(1,1)$ algebra, i.e. $[\hat A_0,\hat A^{(\pm)}]\hat\rho=\pm \hat A^{(\pm)}\hat\rho$, $[\hat A^{(-)},\hat A^{(+)}]\hat\rho=2 \hat A_0\hat\rho$.
This simple discovery, yet of great significance, enables us to carry out the following transformation
\begin{equation}\label{Transformación_1}
    \hat{\rho} = e^{ -\eta \left[\hat{J}_{1}^{(+)} + \hat{J}_{2}^{(+)} \right]} \hat{\varrho}
\end{equation}
in the Lindblad master equation~\eqref{master1}, to obtain
\begin{equation}\label{master2}
\begin{split}
	\frac{d \hat\varrho}{dt}=&-\textrm{i}g\hat S \hat\varrho - G(\eta)\hat\varrho+\sum_{j=1}^2 \gamma_j \left\lbrace \hat J_j^{(-)}(1+ \bar{n}_\text{th})-f(\eta)\hat L_j \right. \\
	& + \left. \left[ 4\eta^2(1+\bar{n}_\text{th})+2\eta(1+2\bar{n}_\text{th})+\bar{n}_\text{th}\right]\hat J_j^{(+)} \right\rbrace\hat\varrho,
	\end{split}
\end{equation}
where $G(\eta)=2(\gamma_1+\gamma_2)[\bar{n}_\text{th} +2\eta (1+\bar{n}_\text{th})]$, and $f(\eta)=(4\eta+1)(1+\bar{n}_\text{th})+\bar{n}_\text{th}$. We can observe that imposing the condition
\begin{equation}\label{quadratic}
4\eta^2(1+\bar{n}_\text{th})+2\eta(1+2\bar{n}_\text{th})+\bar{n}_\text{th}=0,
\end{equation}
we can eliminate the third term of the sum in \eqref{master2}. Condition \eqref{quadratic} is fulfilled when
\begin{equation}\label{eta}
\eta=
\begin{cases}
\displaystyle\eta_{+} = -\frac{\bar{n}_{\text{th}}}{2(1+\bar{n}_{\text{th}})},\\[10pt] 
\displaystyle\eta_{-} = -\frac{1}{2},
\end{cases}
\end{equation}
such that $f(\eta_\pm) = \pm 1$, and
\begin{equation}
G(\eta_\pm)=
\begin{cases}
\displaystyle  0,  &\mbox{if} \quad \eta=\eta_+,\\[6pt] 
\displaystyle -2(\gamma_1+\gamma_2), &\mbox{if} \quad \eta=\eta_-.
\end{cases}
\end{equation}
Without loss of generality, we choose $\eta=\eta_+$, and then the Eq.~\eqref{master2} can be rewritten as
\begin{equation}\label{master2a}
\frac{d \hat\varrho}{dt}= -\textrm{i}g\hat S \hat\varrho +\sum_{j=1}^2 \gamma_j \left[ \hat J_j^{(-)}(1+ \bar{n}_\text{th}) - \hat L_j \right] \hat\varrho.
\end{equation}
Now, by performing the transformation
\begin{equation}\label{Transformación_2}
\hat{\varrho} = e^{-\chi \left[\hat{J}_{1}^{(-)} + \hat{J}_{2}^{(-)} \right]} \hat{\Omega},
\end{equation}
we arrive at
\begin{equation}\label{master3}
\frac{d \hat\Omega}{dt}= -\textrm{i}g\hat S \hat\Omega +\sum_{j=1}^2 \gamma_j\left[\hat J_j^{(-)}(1+\bar{n}_\text{th} -2 \chi) - \hat{L}_j\right]\hat\Omega.
\end{equation}
By setting
\begin{equation}\label{chi}
\chi = \frac{1+\bar n}{2},
\end{equation}
the Eq.~\eqref{master3} reduces to a von Neumann-like equation
\begin{equation}\label{master4}
\frac{d \hat\Omega}{dt} =  -\textrm{i}\left(\hat H_\text{eff}\hat\Omega-\hat\Omega\hat H_\text{eff}^\dagger\right),
\end{equation}
for an  effective non-Hermitian Hamiltonian 
\begin{equation}\label{effectiveH}
\hat H_\text{eff}=-\textrm{i}\gamma_1\hat a_1^\dagger\hat a_1-\textrm{i}\gamma_2\hat a_2^\dagger\hat a_2+g(\hat a_1\hat a_2^\dagger+\hat a_1^\dagger\hat a_2).
\end{equation}
It is proposed that
\begin{equation}\label{omeg}
\hat\Omega = |\psi\rangle\langle\psi|,
\end{equation}
with
\begin{equation}\label{schro}
\textrm{i}\frac{d}{dt}|\psi\rangle = \hat H_\text{eff} |\psi\rangle\quad\text{and}\quad -\textrm{i}\frac{d}{dt}\langle\psi| = \langle\psi|\hat H_\text{eff}^\dagger,
\end{equation}
 a pair of Schr\"odinger-type equations. Thus, the problem has been reduced to solve the equations in~\eqref{schro}, for the non-Hermitian Hamiltonian~\eqref{effectiveH}.

From the results obtained [Eqs.~\eqref{master4} to \eqref{schro}], we emphasize that the two non-unitary transformations \eqref{Transformación_1} and \eqref{Transformación_2} implemented in this work allow us to rewrite the Lindblad master equation for the interaction of two fields with decay at finite temperature as a von Neumann-like equation with an effective non-Hermitian Hamiltonian. Notably, this Hamiltonian is the same as that found in the case of two quantized fields with decay at zero temperature~\cite{Hernandez_2023}. However, the evolution of the system in this new regime experiences passive dissipation processes and thermal excitations from the environment. Despite these differences, we have demonstrated that a direct equivalence can be established between both approaches through this pair of transformations. This connection highlights the importance of non-unitary transformations in understanding thermal dynamics, representing one of the main contributions of this work.

\section{Finite Temperature Master Equation Solutions}\label{Sec_IV}
In order to diagonalize the effective non-Hermitian Hamiltonian~\eqref{effectiveH}, the following operators are introduced
\begin{equation}
    \begin{split}
\hat N &=\hat a_1^\dagger\hat a_1+\hat a_2^\dagger\hat a_2, \\
\hat J_x &= \frac{1}{2}\left(\hat a_1\hat a_2^\dagger+\hat a_1^\dagger\hat a_2\right), 
\\
\hat J_y &= \frac{\textrm i}{2}\left(\hat a_1^\dagger\hat a_2-\hat a_1\hat a_2^\dagger\right),
\\
\hat J_z&= \frac{1}{2}\left(\hat a_2^\dagger\hat a_2-\hat a_1^\dagger\hat a_1\right),
    \end{split}
\end{equation}
satisfying the corresponding commutation relations $[\hat N,\hat J_k]$, with $k=x,y,z$, and $[\hat J_k,\hat J_l]=\textrm{i}\epsilon_{klm}\hat J_m$. The Hamiltonian (\ref{effectiveH}) it is thus expressed as
\begin{equation}\label{Heff}
    \hat H_\text{eff}=-\textrm{i}\left[\frac{\gamma}{2}\hat N+\Delta\hat J_z\right]+2g\hat J_x,
\end{equation}
where $\gamma=\gamma_1+\gamma_2$, and $\Delta=\gamma_2-\gamma_1$. Now, we make use of the non-unitary transformation $\mathcal{\hat{R}} = e^{\xi\hat{J}_y}$,  with $\xi$ a constant (in general complex) to be determined. Thus, we obtain
\begin{equation}\label{H}
\hat{\mathcal{H}} = \mathcal{\hat{R}}^{-1}\hat H_{\textrm{eff}} \mathcal{\hat{R}}= c_N\hat N+c_x\hat J_x+ c_z\hat J_z,
\end{equation}
where $c_N=-\textrm{i}\frac{\gamma}{2}$, $c_x=2g\cosh(\xi)-\Delta\sinh(\xi)$, and $c_z=\textrm{i} \left[ 2g\sinh(\xi)-\Delta\cosh(\xi)\right]$. By setting 
\begin{equation}\label{xi}
  \tanh(\xi)= \frac{2g}{\Delta},\quad\xi= \zeta +\textrm{i}\frac{\tau\pi}{2}, \quad \zeta\in\mathbb R, \quad \tau=0,1,\dots,
\end{equation}
the operator $\hat{\mathcal{H}}$ diagonalizes  as follows
\begin{equation}\label{cases}
\hat{\mathcal{H}}=
\begin{cases}
\displaystyle-\textrm{i}\left[\frac{\gamma}{2}\hat N\pm\sqrt{\Delta^2-4g^2}\hat J_z\right], \quad \mbox{if} & |\Delta|>2g,\\[8pt] 
\displaystyle-\textrm{i}\frac{\gamma}{2}\hat N \mp\sqrt{4g^2-\Delta^2}\hat J_z, \qquad \; \mbox{if} & |\Delta|<2g.
\end{cases}
\end{equation}

Finally, from equations \eqref{Transformación_1}, \eqref{Transformación_2}, \eqref{omeg}, \eqref{schro} and the first equality in~\eqref{H} the exact solution of the Lindblad master equation~\eqref{master1}, given an initial condition $\hat\rho(0)$, is 
\begin{equation}
    \begin{split}
        \hat\rho(t)&
        =e^{ -\eta_+ \left[\hat{J}_{1}^{(+)} + \hat{J}_{2}^{(+)} \right]}e^{-\chi \left[\hat{J}_{1}^{(-)} + \hat{J}_{2}^{(-)} \right] }\hat U(t)\\
        & \quad \times  \left\{e^{\chi \left[\hat{J}_{1}^{(-)} + \hat{J}_{2}^{(-)} \right] }e^{ \eta_+ \left[\hat{J}_{1}^{(+)} + \hat{J}_{2}^{(+)} \right]}\hat{\rho}(0)\right\} \hat U^\dagger(t),
    \end{split}
\end{equation}
where $\chi$ is given in~\eqref{chi}, and
\begin{equation}\label{Operador de Evolucion}
    \hat{U}(t)= \exp\left[\xi\hat{J}_y\right]\exp\left[-\mathrm{i}\hat{\mathcal{H}}t\right]\exp\left[-\xi\hat{J}_y\right],
\end{equation}
is the non-unitary evolution operator associated with $\hat{H}_{\textrm{eff}}$ and $\xi$ is defined by~\eqref{xi}.

\subsection{Example A} \label{Example A}

\begin{figure}[h]
\includegraphics[width= \linewidth]{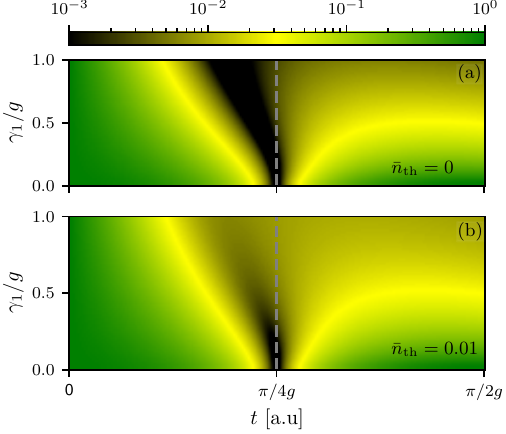}
\caption{Coincidence rates as a function of propagation time $t$ and the decay rate $\gamma_1/g$ (with $\gamma_2 = 0$). In (a), the case at zero temperature ($\bar{n}_{\text{th}} = 0$) is shown, while (b) presents the case at finite temperature with $\bar{n}_{\text{th}} = 0.01$.}
\label{Fig2}
\end{figure}

To validate our analytical results, we investigate the quantum interference of two indistinguishable photons in a cavity with finite-temperature decay. Specifically, we consider two indistinguishable photons in the state $\ket{1,1}$ and calculate the coincidence rate as a function of time $t = t_{\text{out}}$. Fig.~\ref{Fig2} presents complementary results of the coincidence rate in relation to the propagation distance and the loss rates $\gamma_1 / g$, while considering $\gamma_2 = 0$. This configuration allows us to compare the behavior of a photon experiencing decay with that of another photon that remains lossless, helping us understand how these differences influence their quantum interaction.

Fig.~\ref{Fig2}(a) illustrates the photon coincidence rate as a function of propagation distance $t$ in a cavity maintained at zero temperature. A clear minimum in the coincidence rate is observed at $t = \pi /4g$, indicating destructive quantum interference, similar to the experimentally observed Hong-Ou-Mandel effect~\cite{Hong-Ou-Mandel}, and consistent with previous findings in~\cite{Hernandez_2023}. As the decay rate $\gamma_1 /g$ increases, the coincidence rate decreases, and photon bunching occurs at shorter distances compared to previously observed values, reflecting the passive dissipation taking place within the cavity.

In contrast, Fig.~\ref{Fig2}(b) illustrates the behavior of the system at a finite temperature   $\bar{n}_{\text{th}} = 0.01$, where thermal excitations significantly influence the dynamics. While a minimum at $t = \pi /4g$ is still present, its depth is less pronounced compared to the zero-temperature case. This reduction suggests that thermal fluctuations partially disrupt quantum interference, increasing the likelihood of photon coincidence detection. As the decay rate $\gamma_1 /g$ increases, the coincidence rate continues to decrease, though less sharply than in the zero-temperature scenario.

From this, we can observe that as the temperature increases, the system transitions from a regime predominantly governed by quantum interference, as shown in Fig.~\ref{Fig2}(a), to one increasingly influenced by thermal noise, illustrated in Fig.~\ref{Fig2}(b). At zero temperature, quantum interference produces a clear minimum in the coincidence rates; however, as thermal fluctuations become more significant, this minimum diminishes, and the coincidence rates rise due to a higher probability of detecting coincident photons. This shift highlights the profound impact of temperature on photon interactions within the cavity, demonstrating how thermal effects gradually disrupt the quantum coherence that initially dominates the dynamics of the system.

\section{Conclusions}\label{Conclusions}
This study explores the Markovian dynamics of a quantum system characterized by the interaction of two quantized fields subjected to finite temperature decay. By employing superoperator techniques, we have reformulated the Lindblad master equation into a von Neumann-like equation with an effective non-Hermitian Hamiltonian, enabling the exact calculation of the time evolution of arbitrary initial states within a fully quantum framework. Additionally, our analysis of photon coincidence rates offers significant insights into the behavior of indistinguishable photons in a lossy cavity. As the temperature increases, the system transitions from a regime dominated by quantum interference to one increasingly influenced by thermal noise. This transition is reflected in the reduced depth of the quantum interference minimum and the increased coincidence rates, highlighting how thermal conditions impact quantum coherence and the dynamics of the system.

\section*{Acknowledgments}
L. Hernández-Sánchez acknowledges the Instituto Nacional de Astrofísica, Óptica y Electrónica (INAOE) for the collaboration scholarship granted and the Consejo Nacional de Humanidades, Ciencias y Tecnologías (CONAHCYT) for the SNI Level III assistantship (CVU No. 736710). In turn, I. A. B.-G. acknowledges the Spanish MCIN with funding from European Union Next Generation EU (PRTRC17.I1) and Consejería de Educación from Junta de Castilla y León through QCAYLE project, as well as grants PID2020-113406GB-I00 MTM funded by AEI/10.13039/501100011033, and RED2022-134301-T, also CONAHCyT (M\'exico) for financial support through project A1-S-24569, and to IPN (M\'exico) for supplementary economical support through the project SIP20232237.


\end{document}